\documentclass[reqno]{amsart} \usepackage{amscd}
\usepackage{epsfig}
\numberwithin{equation}{section}
\newcommand{\be}{\begin{eqnarray}} 
\newcommand{\ee}{\end{eqnarray}}
\newcommand{\field}[1]{\ensuremath{\mathbb{#1}}}

\newcommand{\ZZ}{\field{Z}}

\DeclareMathOperator{\im}{Im}

\begin{document}
\title[Stringy Description of Schwarzschild Black Holes]{Effective Stringy Description of \\ Schwarzschild Black Holes}
\author{Kirill Krasnov} \address{Albert Einstein Institute\\ 14476 Golm/Potsdam\\ Germany} 
\email{krasnov@aei.mpg.de}
\author{Sergey N. Solodukhin} 
\address{Theoretische Physik\\ Ludwig-Maximilians Universitat\\ Theresienstrasse 37\\
    80333 Munich\\ Germany\\
and International University Bremen\\ P.O.Box 750561\\ 28759 Bremen\\ Germany} 
\email{s.solodukhin@iu-bremen.de
}
\begin{abstract} 
We start by pointing out that certain Riemann surfaces appear rather 
naturally in the context of wave equations in
the black hole  background. For a given black hole there are two closely related surfaces. 
One is the Riemann surface of complexified ``tortoise'' 
coordinate. The other Riemann surface appears when the radial wave equation is 
interpreted as the Fuchsian differential equation. 
We study these surfaces in detail for the BTZ and Schwarzschild black holes in four and higher dimensions. 
Topologically, in all cases both surfaces are a sphere with a set of marked points; for BTZ and 4D Schwarzschild black
holes there is 3 marked points. In certain limits the surfaces can be characterized very explicitly. We then show how 
properties of the wave equation (quasi-normal modes) in such limits are encoded in the geometry of the corresponding 
surfaces. In particular, for the Schwarzschild black hole in the high damping limit we describe the Riemann surface
in question and use this to derive the quasi-normal mode frequencies with the $\log 3$ as the real part.
We then argue that the surfaces one finds this way signal an appearance of an effective string.
We propose that a description of this effective string propagating in the black hole background 
can be given in terms of the Liouville theory living on the corresponding Riemann surface. We give such a stringy 
description for the Schwarzschild black hole in the limit of high damping and show that the quasi-normal modes
emerge naturally as the poles in 3-point correlation function in the effective conformal theory.

\end{abstract}
\maketitle
\section{Introduction}
\label{sec:intr}
It has been suggested some time ago by 't Hooft \cite{G'tH} that there is an analogy between string theory 
and black hole (BH) physics: emission and absorption of particles by a black hole is described by an effective string living on the 2d BH horizon.
More recently AdS/CFT correspondence provided other instances of stringy description of black holes. Another glimpse of relation
between strings and BH's comes from the effective near-horizon description proposed recently in \cite{Solod}.
In the present paper we describe yet another relation between strings and black holes.
We show that in certain regimes an effective stringy description
of black hole arises rather naturally and propose certain (yet to be completed) rules
of this description. The starting point in our approach is the BH wave equation and its 
geometric interpretation in terms of two-dimensional Riemann surfaces.

The black hole wave equations is a fascinating object to study. It has been frequently suggested that the wave equation can
serve as a window into the quantum mechanical description of the black hole. Indeed, according to Hawking, a black hole
emits particles with rate:
\be\label{rate1}
\Gamma(w) \sim \frac{\sigma_{abs}(w)}{e^{w/T_H}-1}.
\ee
Here $w$ is the frequency of a quantum emitted, $T_H$ is the Hawking temperature and we have omitted the density of final states factor 
$d^n k/ (2\pi)^n$ that depends on the number $n$ of spacetime dimensions that we are in. The quantity $\sigma_{abs}(w)$ is the
black hole absorption cross-section which also can be interpreted as the absolute value squared of the transmission coefficient for 
quanta emitted from the horizon. 
It is a complicated function of the frequency which accounts for the transmission
of the emitted quanta through the potential outside of the BH. In a quantum mechanical description one
attempts to replace the whole BH spacetime by a quantum system that would interact with the asymptotic radiation
in exactly the same way as the BH does. In other words, one attempt to describe the BH together with the potential outside.
The scattering of quanta by the potential, emission and absorption processes would then all be described as processes
of interaction of the radiation with a quantum system that replaced the hole. Such a quantum mechanical description must
in particular reproduce \eqref{rate} as the rate of spontaneous emission from the quantum system in question. 
There is by now a famous example when this requirement basically determines what
the quantum system must be. The absorption cross-section for a large class of near-extremal black holes was found to be:
\be\label{abs}
\sigma_{abs}(w)\sim \frac{e^{w/T_H}-1}{(e^{w/2T_L}-1)(e^{w/2T_R}-1)}.
\ee
This formula is valid for the range of frequencies in which the near-horizon approximation can be used (small frequencies, low energy).
Here $T_L, T_R$ are the so-called left and right temperatures, see, e.g. \cite{MS} for more detail. The thermal factors in 
\eqref{abs} are assembled exactly in such a way that the resulting rate is:
\be
\label{rate}
\Gamma(w) \sim \frac{1}{(e^{w/2T_L}-1)(e^{w/2T_R}-1)}.
\ee
This is an emission rate from a quantum system containing two types of bosons, each of them at the corresponding temperature.
Such behavior is characteristic of 2D conformal field theories, where $T_L, T_R$ are the temperatures of left and right movers.
This suggests that the quantum mechanical description of the BH is given by a 2D CFT. This description is valid for small
frequencies, that is, in the near-horizon regime. Thus, in this regime a
near-extremal BH behaves like a 2D CFT! The behavior of the BH absorption cross-section suggested what quantum system must
be used to describe the hole. 

Another instance of a possible glimpse into the black hole quantum physics comes from the study of the BH quasi-normal modes.
It has been speculated in \cite{Hod} that the behavior of the QNM spectrum in the limit of high damping can provide
some hints as to the BH horizon area spectrum quantization. The proposal of Hod is intriguing, and has recently attracted
some attention. In particular Neitzke \cite{Neitzke} has calculated the absorption cross-section
for the Schwarzschild black hole (in any dimension) 
in the same high damping limit and found a suggestive result:
\be\label{3}
\sigma_{abs}(w) \sim \frac{e^{w/T_H}-1}{e^{w/T_H}+3}.
\ee
Thus, in the regime of high damping $w\to i\infty$ the black hole emits particles with the rate:
\be\label{dist}
\Gamma(w) \sim \frac{1}{e^{w/T_H}+3}.
\ee
This answer suggests to search for a system with such statistics. So far, however, 
no genuine quantum system with such properties have been found. 

Thus, properties of the black hole wave equation can be quite instrumental in determining what quantum system describes the hole.
In this paper we study wave equations for various black holes and find that certain Riemann surfaces arise very
naturally in this context. We then use these surfaces to propose an effective quantum mechanical description of
black holes in various limits. Our proposal is that the BH transmission coefficient is related to a 
certain Liouville theory n-point function on
the sphere; the number of insertions depends on the topological type of the Riemann surface that corresponds to the BH.
Thus, the description we propose is essentially a stringy one. Our main application is the limit of high damping, 
or equivalently high energy $|w|\to\infty$. In this case the transmission coefficient is related to a certain
Liouville 3-point function. 

The final word of caution is that the description we give is only an effective one. In this
sense it is similar to e.g. an effective low energy description of rotating black holes found in \cite{MS-1}. 
The question of what are the fundamental degrees of freedom of a BH is not addressed in this paper. We hope,
however, that availability of an effective description will one day aid in finding the fundamental theory
describing black holes quantum mechanically.

The paper is organized as follows. We start by reviewing the black hole wave equations in section \ref{sec:wave}. In section
\ref{sec:surf} we analyze the mapping given by the ``tortoise'' coordinate and describe the Riemann surfaces that appear.
Section \ref{sec:limits} deals with various limits of the wave equation. We show how in such limits the wave equation becomes
solvable and find corresponding Riemann surfaces. These surfaces are then used in section \ref{sec:effective} to propose
an effective Liouville theory description.  We conclude with a discussion of the results obtained. 

\section{Black hole wave equation and the ``tortoise'' coordinate}
\label{sec:wave}

In this paper we shall study non-rotating black holes in various dimensions. The metric for such a BH is of the form:
\be
ds^2 = -g(r) dt^2 + \frac{dr^2}{g(r)} + r^2 d^2\Omega.
\ee
Here $d^2\Omega$ is the metric on the unit sphere of the appropriate dimension. The function $g(r)$ vanishes on the BH horizon. 
The wave equation for scalar perturbations of mass $m$ in this background is:
\be
\frac{1}{\sqrt{-{\rm det}}} \partial_i \sqrt{-{\rm det}}\, g^{ij} \partial_j \varphi = m^2 \varphi.
\ee
It takes the form:
\be
\left(-\frac{1}{g(r)} \partial_t^2 + \frac{1}{r^{n-2}} \partial_r r^{n-2} g(r) \partial_r + \frac{1}{r^2} \Delta_\Omega\right)\varphi=
m^2\varphi.
\ee
Here $n$ is the number of spacetime dimensions and $\Delta_\Omega$ is the Laplacian on the unit sphere. Taking the ansatz:
\be\label{ansatz}
\varphi = e^{-iw t} \frac{\psi(r)}{r^{(n-2)/2}} Y_{l M}(\Omega),
\ee
where $l$ is the angular momentum of the spherical harmonic mode, and $M$ is a collective index of an appropriate length, we get
the radial wave equation:
\be\label{wave}
\left( - \partial^2_z + V(r(z)) - w^2\right)\psi(z) = 0.
\ee
We have introduced the ``tortoise'' coordinate:
\be\label{z}
z(r) = \int \frac{dr}{g(r)}.
\ee
This coordinate is defined up to a constant. The potential in \eqref{wave} is given by:
\be
V(r) = g(r)\left( \frac{l(l+n-3)}{r^2} + \frac{n-2}{2} \frac{1}{r^{(n-2)/2}} \partial_r( g(r) r^{(n-4)/2}) + m^2 \right).
\ee

Thus, the radial wave equation is the Schroedinger equation for a particle moving in the potential $V(r(z))$. For non-extremal BH's
the function $g(r)$ has a first order zero at the horizon. Thus, near horizon
\be
z(r) \sim \log(r-1),
\ee
and therefore $z\to -\infty$ as $r\to 1$. This is why the coordinate is called the ``tortoise'': it takes an infinite amount
of the $z$-coordinate to reach a finite in the $r$-coordinate distance to the horizon. For asymptotically flat BH's the metric
at infinity should approach the flat metric. Thus, $g(r)\to 1$ as $r\to \infty$. Therefore, for large values of $r$ the
tortoise coordinate coincides with the radial one: $z\sim r, r\to\infty$. Thus, in the asymptotically flat case 
the effective quantum mechanical problem is one on the real line: $z\in (-\infty, \infty)$. The potential $V(r(z))$ vanishes
at the horizon: $V(z)\to 0, z\to-\infty$ and approaches constant $V(z)\sim m^2$ at infinity. 
For functions $g(r)$ coming from the BH metrics the potential is everywhere positive. For massless fields the
potential is zero at both ends and 
therefore looks like the snake that ate an elephant from Antoine de Saint Ex\'upery's book
``The Little Prince''. Its hight in the maximum depends on $l$. 
The absorption cross-section $\sigma_{abs}$ must be obtained by considering the transmission in this potential. The
quasi-normal modes are poles of $\sigma_{abs}$. 

Apart from the Schroedinger equation the wave equation has yet another useful interpretation. It is that of the Fuchsian
differential equation from the theory of conformal mappings. We turn to this interpretation in the next section.

\section{Riemann surfaces}
\label{sec:surf}

The following differential equation appears prominently in the theory of conformal mappings:
\be\label{fuchs}
\partial_z^2 \psi + \frac{1}{2} {\it T} \psi = 0.
\ee
Here ${\it T}={\it T}(z)$ is a holomorphic function of $z$. The equation \eqref{fuchs} is known as the Fuchsian differential
equation. Its characteristic property is that the ratio of two linearly independent solutions:
\be
w(z):= \psi_1(z) / \psi_2(z)
\ee
is a map from the complex $z$-plane to the complex $w$-plane such that its Schwarzian derivative equals to $\it T$:
\be
{\mathcal S}(w;z) := \frac{w'''}{w'} - \frac{3}{2} \left(\frac{w''}{w'}\right)^2 = {\it T}(z).
\ee
The problem of solving  the differential equation \eqref{fuchs} can be reduced to the problem of finding the image
of the complex $z$-plane under the map $w(z)$: on the Riemann surface, that is the image under this map, the
equation \eqref{fuchs} becomes the trivial equation:
\be
\partial^2_w \psi = 0.
\ee
Thus, one can immediately write down the general solution of \eqref{fuchs} if the map $w(z)$ is known. The problem of solving
the wave equation thus reduces to the problem of finding the corresponding Riemann surface. 

In the present paper we would like to apply this point of view to the black hole wave equations, and describe the corresponding
Riemann surfaces. Since the wave equation becomes trivial on the corresponding surface, the surface itself characterizes the
scattering problem completely and is thus of fundamental importance. Let us start from the simplest example in 
which the wave equation can be solved exactly: the case of the 2+1 dimensional BTZ BH.

\subsection{BTZ BH wave equation}

\begin{figure}
\centering 
\epsfig{figure=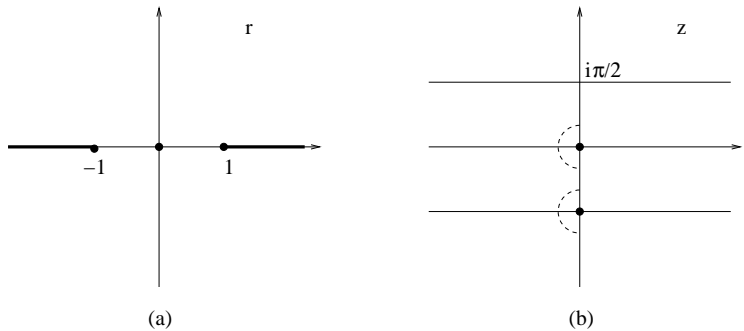, height=1.7in}
\caption{The tortoise map for the BTZ BH. 
(a) The complex $r$-plane is shown together with the branch cut (bold line); (b) The image of the $r$-plane on the $z$-plane
is the strip of width $\pi$. Both horizons are mapped to infinity, zero is mapped to zero, and infinity is mapped to the point $z=-i\pi/2$.
Closed contours around the singularity and infinity are shown.}
\label{fig:btz}
\end{figure}
%
The BTZ BH is not asymptotically flat; the corresponding function $g(r)$ is given by:
\be
g(r) = r^2-1.
\ee
We have again placed the horizon at $r_+=1$. The potential for this BH is given by:
\be
V(r) = (r^2-1)\left( \frac{l^2}{r^2} - \frac{1}{4} \frac{r^2-1}{r^2} + 1 +m^2\right).
\ee
Here $l$ is the angular momentum quantum number on the circle.
Unlike the asymptotically flat case, the potential grows at infinity as $V(r)\sim r^2(3/4+m^2)$.
We note that neither of these functions is sensitive to the sign of $r$. Thus, one must identify
$r\sim -r$. The tortoise coordinate \eqref{z} is given by:
\be\label{z-btz}
z(r) = \frac{1}{2} \log{\frac{r-1}{r+1}} - \frac{i\pi}{2}.
\ee
We have chosen the additive constant so that $z(r)=0$ when $r=0$. This choice will prove convenient for what follows.
Let us now see how the complex $r$-plane is mapped to the $z$-plane. Since $\pm r$ are to be identified, it is
enough to consider only the right half-plane. However, as usual, it is easier to work with the covering space. Let us first give the 
picture of the mapping of the covering space. The logarithm function
requires a branch cut. We shall place the cut along the real axis, from $-\infty$ to $-1$, and from $1$ to $\infty$.
Thus, on the compactified $r$-plane this is simply a cut from $-1$ to $1$ passing through infinity. The $r$-plane
is mapped into the strip $\im{z}\in[-\pi/2,\pi/2]$ in the complex $z$-plane.  
The horizons $r=\pm 1$ are mapped to infinity, and the singularity $r=0$ is mapped to zero. The infinity in the $r$-plane 
is mapped to the point $z=-i\pi/2$. See Fig.\ref{fig:btz}.

\begin{figure}
\centering 
\epsfig{figure=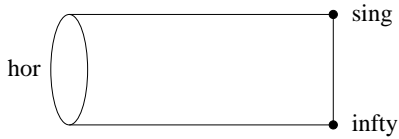, height=0.7in}
\caption{The surface obtained after all the identifications are performed.}
\label{fig:btz-i}
\end{figure}

Thus, if not for the identification $r\sim -r$, the
image of the $r$-plane under the tortoise coordinate map would be a strip with its two edges identified, that is a cylinder. 
Let us now describe the result of
the identification. The action of the identification on the $z$-plane is: $z\sim -z$. The quotient is therefore half 
the cylinder, with one of the end holes sealed. The resulting surface is shown in Fig.\ref{fig:btz-i}. Topologically this 
is a sphere with a hole and two conical singularities of angle deficit $\pi$ each.
%

Thus, we get a wave equation in the complex $z$-plane. In the case at hand the map $z(r)$ is not very complicated; it can
be explicitly inverted: $r=\tanh{(-z)}$. It is then easy to write the wave equation directly in terms of the $z$-coordinate.
One gets an equation solvable in terms of the hypergeometric function. To bring it into the explicitly hypergeometric form
one makes a further change of variables that maps the infinity on the $z$-plane to zero, and the point
$z=i\pi/2$ to one. Instead of describing this change of coordinates let us follow another route. Let us directly get the Riemann surface that
``solves'' the wave equation. We shall use the following reasoning. The wave equation is one on the surface shown in Fig.\ref{fig:btz-i}.
The potential has 3 singular points: horizon, singularity and the infinity. As one can easily check, the leading singularity at
these points is that of a double pole. This
is suggestive of the Fuchsian equation that corresponds to the hypergeometric function; we review this equation in the
appendix. Knowing that the equation is of the hypergeometric type one can find the parameters of the hypergeometric
function by calculating the monodromy around the singular points. This way one determines the parameters directly, without 
having to do a coordinate transformation.

Thus, we would like to calculate the monodromy of the BTZ BH wave equation around the horizon, singularity and infinity.

\bigskip
\noindent{\bf Horizon.} At the horizon the potential vanishes, the wave equation becomes:
\be\label{eq-hor}
(\partial_z^2 + w^2)\psi = 0.
\ee
Its two linearly independent solutions are: $\psi_{1,2} = e^{\pm iwz}$. A rotation by $2\pi$ around the horizon  is a
shift $z\to z+i\pi$. The two solutions transform: $\psi_1 \to e^{-\pi w} \psi_1, \psi_2\to e^{\pi w} \psi_2$. 
The corresponding monodromy matrix is ${\bf M}_h = {\rm diag}(e^{-\pi w}, e^{\pi w})$. The trace is equal to:
\be
\frac{1}{2}{\rm Tr} {\bf M}_h = \cosh{\pi w}.
\ee

\bigskip
\noindent{\bf Singularity.} The tortoise coordinate map $z(r)$ behaves near the singularity as: $z=-r+O(r^2)$. Thus, keeping only the leading 
term of the potential, the equation becomes:
\be
\left(\partial_z^2 + \frac{1+4l^2}{4 z^2} + w^2\right)=0
\ee
One gets the equation solvable in terms of the Bessel functions. The two linearly independent solutions are:
\be
\psi_{1,2} = \sqrt{z} J_{\pm il}(wz).
\ee
A tricky point is that the closed contour around the singularity is a rotation by $\pi$, not by $2\pi$. This is
due to the identification $z\sim -z$ that must be performed on the $z$-plane. Thus,
a rotation around $z=0$ by $\pi$ results in: $\psi_1 \to e^{i\pi/2-\pi l}\psi_1, \psi_2\to e^{i\pi/2+\pi l}\psi_2$.
Thus, the corresponding monodromy matrix is: ${\rm diag}(ie^{-\pi l},ie^{\pi l})$. As it stands, its determinant
is not equal to one. We have to divide this matrix by the square root of its determinant. The correctly normalized 
monodromy matrix is given by: ${\bf M}_s = {\rm diag}(e^{-\pi l},e^{\pi l})$; its trace is
given by:
\be
\frac{1}{2}{\rm Tr}{\bf M}_s = \cosh(\pi l).
\ee

\bigskip
\noindent{\bf Infinity.} Infinity in the complex $r$-plane is mapped to $z=-i\pi/2$. Introducing a new coordinate
$z=-i\pi/2+y$, and again keeping only the leading term of the potential $V(y)=y^2(3/4+m^2)$, we get an equation:
\be
\left(\partial_y^2 - \frac{3+4m^2}{4 y^2} + w^2\right)=0.
\ee
It is solution is given in terms of the Bessel function:
\be
\psi_{1}=\sqrt{y} J_\nu(wy), \psi_2 = \sqrt{y} J_{-\nu}(wy).
\ee
Here 
\be
\nu=\sqrt{1+m^2}.
\ee
We have assumed that $\nu$ is not an integer and one can use $J_{\pm\nu}$ as two linearly independent solutions. If this is not
so, one has to use the Bessel function of the second kind as the second linearly independent solution. In any case, a closed 
contour around infinity is a rotation by $\pi$ around $y=0$. The monodromy matrix is calculated using \eqref{bessel-mon}. One gets:
\be
\frac{1}{2}{\rm Tr}{\bf M}_\infty = \cos{\pi\nu}.
\ee
This result holds independently of whether $\nu$ is an integer or not.

\begin{figure}
\centering 
\epsfig{figure=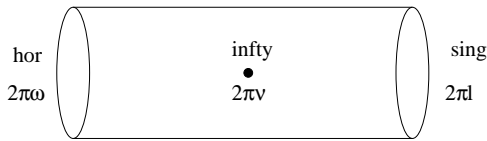, height=0.7in}
\caption{The W-surface for the BTZ BH}
\label{fig:btz-w}
\end{figure}

Having calculated the monodromies of the wave equation, it is easy to write the hypergeometric equation with the same
monodromies. One uses the fact that the monodromies around the singular points of the hypergeometric equation 
are given by $\cos(\pi a_i)$. One immediately gets:
\be\label{index-btz}
a_1=iw, \qquad a_2=\nu, \qquad a_3=il.
\ee
Notice that $a_1=2\pi T_H i \omega$ in terms of the Hawking temperature $T_H=1/2\pi$. Written in this form the 
expression for the monodromy around the horizon is also valid for the Schwarzschild black hole in any dimension, as 
we show later in the paper.
The quantities \eqref{index-btz} are correct parameters that one can obtain, for example, by doing a coordinate transformation that
brings the wave equation into the hypergeometric form, see \cite{Cardoso,Danny}. The resulting surface that solves the wave equation
is shown in Fig.\ref{fig:btz-w}.

Thus, a solution of the resulting wave equation is the
hypergeometric function $F(\alpha,\beta;\gamma ;y)$ with singular points
at $y=0$ (singularity), $y=1$ (infinity) and $y=\infty$ (horizon) and parameters:
\be\label{index-btz-a}
\alpha=\frac{1}{2}(1-\nu+i(l-w)), \qquad \beta=\frac{1}{2}(1-\nu-i(l+w)), \qquad \gamma=1-iw.
\ee
as determined in terms of  monodromies \eqref{index-btz} by \eqref{parameters}.
%

Having determined the parameters \eqref{index-btz}, \eqref{index-btz-a} it is quite easy to determine the quasi-normal modes. 
One has to impose the following
boundary conditions: the solution should vanish at infinity and be in-going at the horizon. It is possible to satisfy these
boundary conditions only at a discrete set of frequencies: the quasi-normal modes. As the solution that is purely in-going at the
horizon one takes, see \cite{Danny}, the hypergeometric function $F(\alpha,\beta;\gamma;y)$ 
multiplied by $y$ to certain power. Here $y$ is a coordinate that
is different from the tortoise coordinate. It is the one in which the wave equation has an explicitly hypergeometric form. The
horizon is located at $y=0$ in this coordinates. The relation between $y$ and $z$ is $y=-1/\sinh^2{z}$. Note that 
our method works with monodromies only and we never need to use this relation. This solution can be expanded into two
linearly independent solutions near $y=1$ (infinity). One has to use the formula \eqref{app:h} given in the Appendix.
The second term in proportional to $(1-y)$ and vanishes at $y=1$. The condition that the first term vanishes is
satisfied at the poles of the $\Gamma$ functions in the denominator. We get two sets of poles:
\be
\frac{1}{2}+\frac{\nu}{2}-\frac{iw}{2} -\frac{il}{2}=-n, \qquad \frac{1}{2}+\frac{\nu}{2}-\frac{iw}{2} +\frac{il}{2}=-n.
\ee
In terms of the monodromies (\ref{index-btz}) these two sets are represented as follows
\be
1+a_2-a_1-a_3=-2n~,~~1+a_2-a_1+a_3=-2n~~.
\label{aaa}
\ee
This gives the correct, see \cite{Cardoso,Danny}, quasi-normal modes. Recalling that the Hawking temperature for the BTZ (in the units
when $r_+=1$) is equal to $T_H=1/2\pi$ the frequencies can be rewritten in a more familiar form:
\be\label{qnm-btz}
w = \pm l - 4\pi i T_H (n+h),
\ee
where we have introduced:
\be
h=\frac{1}{2}(1+\nu)=\frac{1}{2}(1+\sqrt{1+m^2}).
\ee
This is usual expression for the conformal dimension of a field of mass $m$ in AdS/CFT correspondence.

We would like to point out that there is a quicker root to the quasi-normal modes. Let us consider the case $m=0$. Then,
quite remarkably, the QNM condition can be expressed geometrically as the condition that the sizes of the two holes of the $W$-surface
are equal: 
\be
\cos{\pi a_1}=\cos{\pi a_3}.
\label{cos}
\ee
This gives the correct expression $w=\pm l-2ni$. We shall see how a similar geometric condition leads to the quasi-normal
modes for the Schwarzschild BH. Equivalently, the equation (\ref{cos}) can be written in the form
$$
e^{i\pi a_1} \ e^{i\pi a_3}=1~~.
$$
This is the form which can be further generalized to include  the case of non-vanishing mass 
$m$ when the monodromy 
$a_2$ around the infinity is non-trivial. The generalization can be read from the condition
(\ref{aaa}) and takes the form
\be
e^{i\pi a_1} \ e^{-i\pi a_2} \ e^{i\pi a_3}=e^{i\pi}~~.
\label{aaap}
\ee
We  show later in the paper that the condition (\ref{aaa}) and respectively (\ref{aaap})
emerge naturally as a condition for location of poles in the Liouville 3-point function.
The condition similar to \eqref{aaap} was suggested in \cite{BC} for monodromies around inner and outer
horizons of the rotating BTZ black hole.

Thus, we have learned that the wave equation has two closely related Riemann surfaces associated with it. One surface, which
we shall refer to as the $Z$-surface, is simply the image of the $r$-plane under the tortoise coordinate map $z(r)$.
In the case of the BTZ BH this surface is shown in Fig.\ref{fig:btz-i}. The BH wave equation is the Fuchsian
equation on this surface. In the case of the BTZ BH this is the hypergeometric equation with the parameters
given by \eqref{index-btz}. The ratio $w(z)=\psi_1/\psi_2$ of two linearly independent solutions of the hypergeometric 
equation maps the complex $z$-plane into a sphere with either holes
or conical singularities. In our case this is a sphere with two holes of sizes $2\pi w$ and $2\pi l$, and a 
single ``conical defect'' whose total angle at the tip is $2\pi\nu$.
We shall refer to the Riemann surface that is obtained as the result of the $w$-map as the $W$-surface. Topologically,
both $Z$ and $W$-surfaces are spheres with 3 marked points. The $Z$-surface codes only the properties of the
tortoise coordinate map. It does not know anything about the potential. The wave equation is an equation on the $Z$-surface. 
The $W$-surface ``solves'' this wave equation in the sense that if one knows the map $w(z)$ and its inverse one
can explicitly write down the solution on the $Z$-surface. The $W$-surface knows about the potential. Its moduli
depend on the parameters of the potential, which are the parameters of the BH. It is the $W$-surface that is
the main object of interest. However, as the example of the BTZ BH shows, the $Z$-surface is of the same
topological type. It is therefore of interest as well.

Let us now show how considerations of this subsection carry over to more interesting black holes.

\subsection{Schwarzschild BH wave equation}

\begin{figure}
\centering 
\epsfig{figure=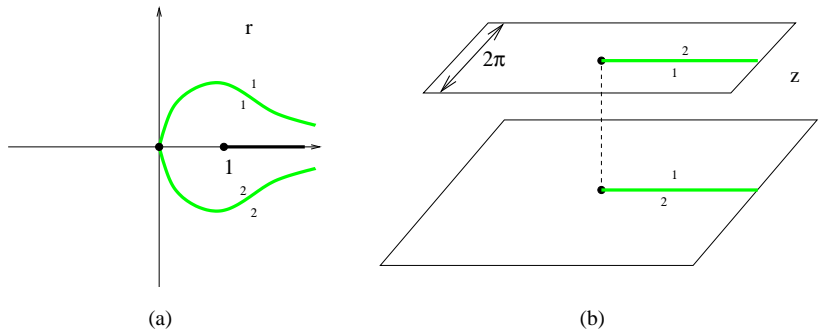, height=1.7in}
\caption{The tortoise map for the 4d Schwarzschild BH. 
(a) The complex $r$-plane with a cut placed from the horizon to infinity. The region inside the green curve is mapped into
the physical sheet (the strip). The green curve itself becomes the cut on the $z$-plane 
(b) The $Z$-surface has two sheets. The physical sheet is the strip of width $2\pi$; the unphysical
sheet is a copy of the complex plane. The sheets are glued as indicated.}
\label{fig:4}
\end{figure}
%

For the Schwarzschild BH in $n=4$ spacetime dimensions the function $g(r)$ is given by (we put the horizon at
$r_+=1$):
\be
g(r) = 1-\frac{1}{r},
\ee
and one gets the familiar potential:
\be\label{pot-4d}
V(r) = \left(1-\frac{1}{r}\right) \left( \frac{l(l+1)}{r^2} + \frac{1}{r^3}\right).
\ee
The tortoise coordinate \eqref{z} is given by:
\be\label{z-4}
z(r) = r + \log{(r-1)} - i\pi.
\ee
Again, we have chosen the additive constant so that the singularity $r=0$ is mapped to zero. The map \eqref{z-4} is
now much more involved than in the case of the BTZ BH. To understand the image of the complex $r$-plane it
is instructive to consider what the circles $r=1+ Re^{i\phi}$ are mapped to. The presence of the logarithm requires
a branch cut. We put it along the positive real axis from the horizon $r=1$ to infinity. Then small circles
$R<<1$ are mapped into almost vertical segments of size $2\pi$. The image of the circle gets more complicated as
$R$ increases. When $R=1$ the image develops a kink. This kink is located exactly at the singularity $z=0$.
For larger radii the image of the circle has a self-intersection. The self-intersection is located on the
positive $z$-axis. A moment of reflection reveals the structure shown in Fig.\ref{fig:4}. One has a strip of
width $2\pi$; with a cut along the positive real axis. The sides of the strip are to be identified. 
Below the strip one attaches a copy of the complex plane. It is also cut along the positive real axis. 
One identifies the sides of the cuts as
is shown in Fig.\ref{fig:4}. Note that the rotation by $2\pi$ around the singularity is now 
a rotation by $4\pi$ in the $z$-plane. This can be seen by expanding the logarithm near $r=0$. One finds:
$z=-r^2/2 +O(r^3)$. Half of this rotation occurs on the upper sheet, the other half is on the lower sheet.
We shall refer to the two sheets in Fig.\ref{fig:4} as the physical (upper) and unphysical (lower).
The two sheets glued together form the $Z$-surface for the Schwarzschild BH.

\begin{figure}
\centering 
\epsfig{figure=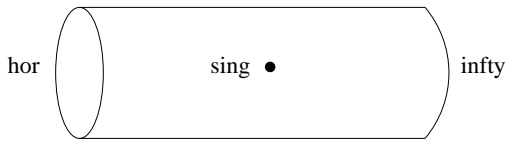, height=0.7in}
\caption{The topological type of the Schwarzschild BH $Z$-surface is that of a cylinder with one marked point. The marked point corresponds 
to the singularity.}
\label{fig:4-i}
\end{figure}

Let us now discuss the topological type of the $Z$-surface. The physical sheet is simply a cylinder. One attaches to it
a copy of the complex plane. However, this does not change the topological type. Thus, topologically, the $Z$-surface
is simply a cylinder with an additional marked point- the singularity. Let us note however that the two ends of the cylinder 
will have different monodromy properties when we consider the wave equation on the $Z$-surface. Thus, the ends of the cylinder 
are different. One of them corresponds to the horizon, the other is the infinity. The corresponding surface
is shown in Fig.\ref{fig:4-i}.
%

Let us now discuss the wave equation on the $Z$-surface. Unfortunately, now the potential is too complicated to be
solved exactly. Thus, even though the $Z$-surface is still topologically a sphere with 3 marked points
as in the case of BTZ black hole, no
exact solution in terms of the hypergeometric function is possible. However, the hypergeometric equation
can be used as a good approximation in certain regimes. Such approximations are studied in the next section.
Let us now turn to the higher
dimensional Schwarzschild black holes, and see what kind of surfaces one gets.

\subsection{Higher-dimensional Schwarzschild BH wave equations}

The function $g(r)$ characterizing a BH in higher dimensions is given by:
\be
g(r) = 1-\frac{1}{r^{n-3}}.
\ee
As before, the horizon is at $r_+=1$. The potential that one gets is:
\be
V(r)=\left(1-\frac{1}{r^{n-3}}\right)\left( \frac{l(l+n-3)}{r^2} + \frac{n-2}{4}\left(\frac{n-4}{r^2}+\frac{n-2}{r^{n-1}}\right) \right).
\ee
The tortoise coordinate is given by:
\be
z(r) = r+\sum_{j=0}^{n-4} \frac{e^{2\pi i j/(n-3)}}{n-3} \log{(1 - re^{-2\pi i j/(n-3)})}.
\ee

\begin{figure}
\centering 
\epsfig{figure=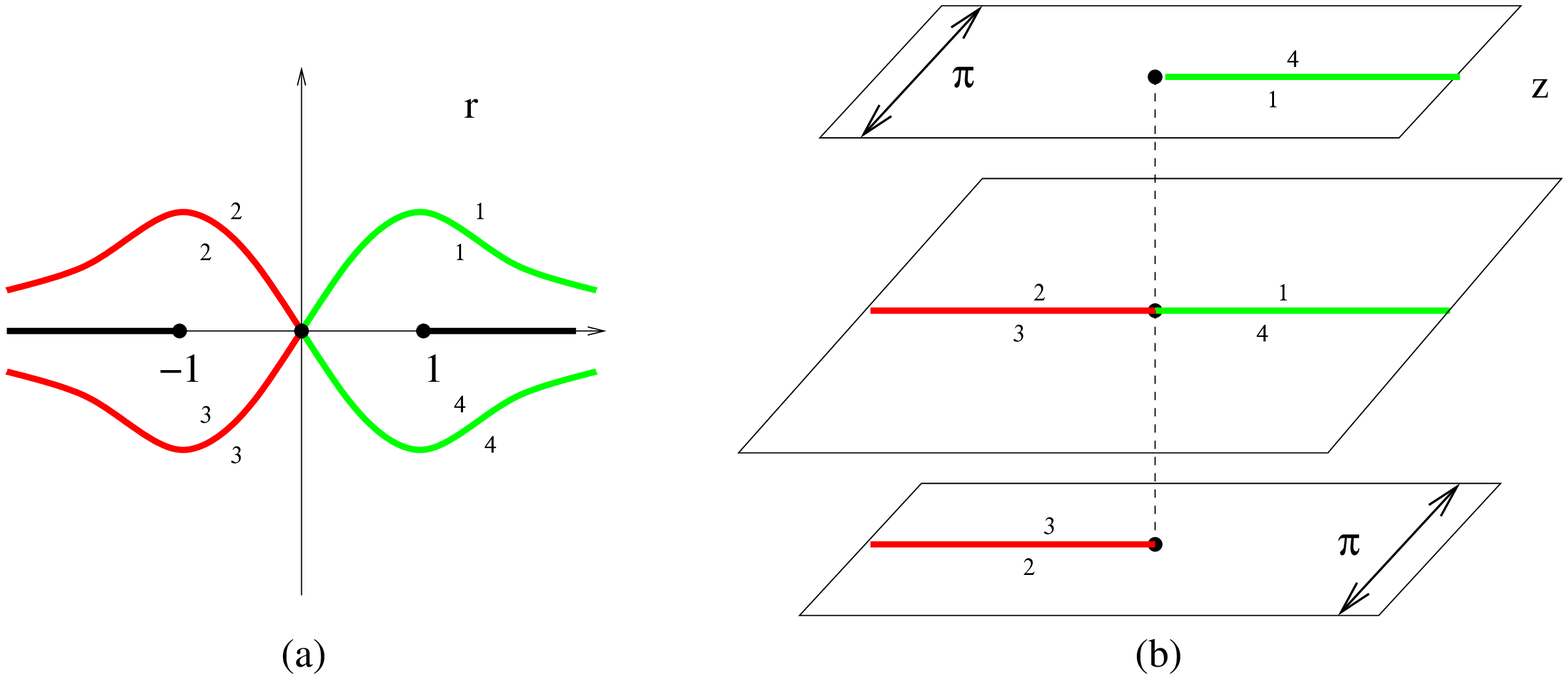, height=1.8in}
\caption{The tortoise map for the 5d Schwarzschild BH. 
(a) The complex $r$-plane has now a cut placed between the horizons going through infinity. The green and red curves are mapped 
to the cuts in the $z$-plane; (b) The $Z$-surface of the 5-dimensional Schwarzschild BH consists of 3 sheets. Two of the sheets are strips of
width $\pi$, the other sheets is a copy of the complex plane. The gluing identifications are indicated.}
\label{fig:5}
\end{figure}
%
As before, we would first like to understand the $Z$-surface that appears as the result of the tortoise coordinate mapping. Let us consider the 
case $n=5$; the general pattern will then become clear. In this case the tortoise map becomes:
\be
z(r) = r+ \frac{1}{2} \log{\frac{1-r}{1+r}}.
\ee
This is ``almost'' the BTZ BH tortoise coordinate \eqref{z-btz}, if not for the first term. It is the presence of this term that
makes the Riemann surface very involved. As in the BTZ case, the presence of the logarithm requires a branch cut. We place it from
the horizon at $r=1$ to the other, unphysical horizon at $r=-1$ along the real axis and avoiding the singularity. Thus, the cut goes through
the infinity. Near the singularity the tortoise coordinate behaves as:
\be
z=-\frac{r^3}{3}+O(r^{4}).
\ee
Thus, the singularity $r=0$ is mapped to $z=0$, and near the singularity the tortoise map is a $1$ to $3$ covering. One should expect
that the relevant Riemann surface will contain 3 sheets. The surface is shown in Fig.\ref{fig:5}.

\begin{figure}
\centering 
\epsfig{figure=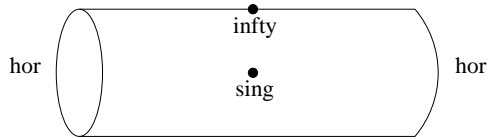, height=0.7in}
\caption{The topological type of the 5d Schwarzschild BH $Z$-surface is that of a cylinder with 2 marked points.}
\label{fig:5-i}
\end{figure}

The topological type of the $Z$-surface can be determined by thinking about the geometry in Fig.\ref{fig:5}(b). It is a cylinder, with two ends 
corresponding to the two horizons, and a handle with one of the cycles completely degenerated. The degenerated cycle is at 
infinity. We believe that it is essentially right to replace the degenerated handle by a marked point, denoted as infinity
in Fig.\ref{fig:5-i}. The other marked point is the singularity.

Being topologically a sphere with 4 marked points the surface is too complicated to approximate the corresponding $W$-surface by
the hypergeometric equation. Thus, no analysis similar to the one performed in the BTZ case is possible. However, 
simplifications are possible in certain limits, as we shall see in the next section. 

Let us now turn to the discussion
of $Z$-surfaces for Schwarzschild black holes in higher dimensions. 
Expanding the tortoise map near the singularity one finds, in general:
\be
z(r) = -\frac{r^{n-2}}{n-2} + O(r^{n-1}).
\ee
Thus, the map near the singularity is $1$ to $n-2$. Therefore, the $Z$-surface consists of $n-2$ sheets. One of these sheets is the
physical sheets containing the horizon. It is always a strip of width $2\pi/(n-3)$. Topologically,
the $Z$-surface is a sphere with $n-1$ marked points. Out of these points $n-3$ are holes corresponding to the horizons. The other
marked points are the singularity and the infinity. For $n$ even infinity is a hole, for $n$ odd it is just a marked point on 
the surface.

%

The wave equations are too complicated to be treated exactly. However, since the singularities of the potential coincide
with the singularities (marked points) of the $Z$-surface, one does not expect that the map $w(z)$ will change the
type of the Riemann surface. Thus, it is to be expected that the $W$-surface that ``solves'' the wave equation
is of the same topological type as the $Z$-surface. Since the potential is coded in the $W$-surface, so are 
the BH parameters. 

\section{Limits in which the wave equation becomes solvable}
\label{sec:limits}

In this section we will apply the techniques described to solve the wave equation in certain limits. In order to obtain a solvable
wave equation one has to approximate the potential by a solvable one. In practice this is done by keeping only 
the leading term of the potential in some limit, and ignoring the other terms. One can classify such approximations 
according to where on the $r$-plane the term kept is the leading singularity. Thus, there is an approximation
that is good near the horizon or near singularity. Each of this approximations leads to a different
wave equation. 

Another, more geometrical point of view on this is as follows. As we have explained,
the wave equation is ``solved'' by a particular $W$-surface. Thus, an approximate solution is obtained by approximating
the real $W$-surface with another one, which can be described explicitly. Approximating the potential by its behavior
near one of the singular points on the $Z$-surface, one essentially zooms onto the region close to this singular point.
Thus, the $W$-surface that arises as the result of the approximation is a part of the ``correct'' $W$-surface
near one of the singularities. Such an approximation is not good for all frequencies. However, in certain limits,
the approximated $W$-surface captures all physics.

The main limit
that we are interested in is that of high damping $w\to i\infty$. However, it is instructive to describe the other possible 
limits as well. We shall consider two interesting limits:
\begin{itemize}
\item The limit of low energy $|w|\to 0$. This corresponds to keeping only the terms most relevant near the horizon of the BH.
\item The limit of high energy $|w|\to \infty$. In this limit the relevant region is that where the potential becomes
large. One keeps only the leading terms near the singularity.
\end{itemize}
We restrict our analysis to the case of the 4d Schwarzschild BH; some comments on the general case will be given after this analysis.
Let us first consider the low energy limit. It has been analyzed in detail in \cite{Solod} so we shall be very sketchy.

\bigskip
\subsection{\bf Low energy limit.} 
In this approximation it is convenient to define the tortoise coordinate a bit differently:
\be
\tilde{z} = r-1 + \log{(r-1)}.
\ee
This amounts to a different choice of the integration constant in \eqref{z}. 
The potential for the 4d Schwarzschild BH is given by \eqref{pot-4d}. Near the horizon it can be approximated by:
\be\label{pot-low}
V(z) = \delta^2 e^{\tilde{z}} - \lambda^2 e^{2\tilde{z}}, 
\ee
where $\delta^2= l(l+1)+m^2+1, \lambda=4l(l+1)+2m^2+5$. The expansion whose first two terms are \eqref{pot-low} is
in powers of $(w/l)^2$. Thus, it is good when $|w|<< l$. Near the horizon $e^{\tilde{z}}=r-1 + O((r-1)^2)$.
In the near-horizon approximation one keeps only the first term. Then $\tilde{z}=\log(r-1)$. The corresponding
$Z$-surface is a cylinder of circumference $2\pi$. Restricting potential \eqref{pot-low} to the first term
we get a classical Liouville equation while the second term gives a perturbation.

The wave equation with \eqref{pot-low} can be solved in terms of confluent hypergeometric functions. Let us introduce:
$\phi=e^{-\tilde{z}/2}W$ and a new variable $y=2i\lambda e^{\tilde{z}}$. The wave equation becomes:
\be
\frac{d^2 W}{dy^2} + \left( -\frac{1}{4}+\frac{i\delta^2}{2\lambda y} +\frac{\frac{1}{4}+w^2}{y^2}\right) W =0.
\ee
This is the confluent hypergeometric equation in the form \eqref{whitt}. Thus, two linearly independent solutions of the
original wave equation are:
\be\label{sol-low}
\phi_1 = e^{-\tilde{z}/2} W_{\frac{i\delta^2}{2\lambda},iw}(2i\lambda e^{\tilde{z}}), \qquad
\phi_2 = e^{-\tilde{z}/2} W_{-\frac{i\delta^2}{2\lambda},iw}(-2i\lambda e^{\tilde{z}}).
\ee
It is now easy to evaluate the monodromies. 

\bigskip
\noindent{\bf Horizon.} Near the horizon the potential vanishes. The wave equation becomes \eqref{eq-hor}. The monodromy calculation is then 
unchanged from the BTZ case. The only difference is
that now the size of the strip is $2\pi$ instead of $\pi$. This results in the monodromy:
\be\label{mon-h-low}
\frac{1}{2}{\rm Tr}{\bf M}_h = \cosh{2\pi w}.
\ee

\bigskip
\noindent{\bf Infinity.} The asymptotic behavior of the Whittaker's functions is given by \eqref{whitt-as}. This results in the 
following monodromy:
\be
\label{mon-infty-low}
\frac{1}{2}{\rm Tr}{\bf M}_\infty = -\cosh{\pi\delta^2/\lambda}.
\ee
Notice an additional minus sign in front of the $\cosh$ that comes from the factor $e^{-\tilde{z}/2}$ multiplying 
Whittaker's functions in \eqref{sol-low}. 

The resulting $W$-surface is now a cylinder with sizes of holes $4\pi w$ and $2\pi\delta^2/\lambda$. As in the BTZ case, the quasi-normal 
modes can be determined from a geometric condition saying that the sizes of holes are equal. 
Thus equating monodromies (\ref{mon-h-low}) and (\ref{mon-infty-low}) this gives the correct, 
see \cite{Solod}, expression for the 
QNM frequency in the potential \eqref{pot-low}: 
\be\label{qnml}
w=2\pi T_H\delta^2/\lambda-4\pi i T_H (n+\frac{1}{2})~~.
\ee

\bigskip
\subsection{\bf High energy limit.} 

\begin{figure}
\centering 
\epsfig{figure=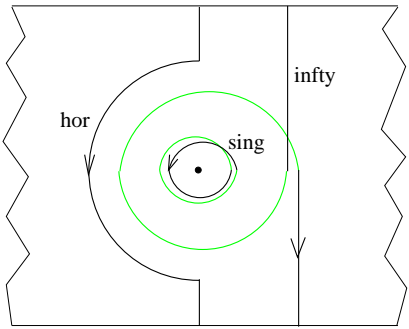, height=1.7in}
\caption{Contours for evaluation of monodromies in the case of Schwarzschild black hole high energy limit. Parts of contours shown in 
green are on the unphysical sheet.}
\label{fig:cont}
\end{figure}
%

As it was demonstrated in \cite{Motl}, in the limit $|w|\to\infty$ it is correct to keep only 
the leading singularity of the
potential near $r=0$. For the 4d Schwarzschild, this singularity is $V(r) \sim -1/r^4 \sim -1/4z^2$. It can be shown, see \cite{Siopsis},
that the sub-leading term is
$-[l(l+1)+1/3]/(-2z)^{3/2}$. One can rewrite the wave equation in terms of the rescaled coordinate $y=wz$. Then 
the sub-leading term will be of order $[l(l+1)+1/3]/w^{1/2}$, and can be dropped in the large $|w|$ limit. Thus, the high energy limit
is that of $|w|>> l^4$. In this approximation one has to use:
\be\label{pot-4-a}
V(z) = -\frac{1}{4z^2}
\ee
{\it for all} values of $z$, not necessarily small. Actually, one has to be a bit careful here because the potential must be
periodic on the physical sheet (the strip) and \eqref{pot-4-a} is not. Thus, to be more precise, we approximate the potential 
by \eqref{pot-4-a} only on the unphysical sheet. 

Having spelled out the approximation that we are going to use, it is easy to determine a solvable wave equation with 
the same properties. Indeed, the $Z$-surface in question is a sphere with 3 marked points. A solvable wave
equation in this case must be of the hypergeometric type. It is therefore completely determined by its
monodromies. Thus, we only have to compute the monodromies around the singularities.

Thus, the potential in this approximation is given by \eqref{pot-4-a}.
The wave equation is of the Bessel type. Two linearly independent solutions are given by
$\psi_1=\sqrt{z} J_0(wz), \psi_2=\sqrt{z} N_0(wz)$. The contours for evaluating monodromies are
shown in Fig.\ref{fig:cont}. The calculation is as follows.

\bigskip
\noindent{\bf Horizon.} In the high energy approximation one zooms onto the singularity region. Thus, the contour around the
singular point that corresponds to the horizon looks as shown in Fig.\ref{fig:cont}. In particular,
this closed contour involves not only the shift $z\to z+2\pi i$, but also a rotation around singularity 
by $\pi$ in the anti clock-wise direction. Thus, one must first 
perform the shift $z\to z+i\pi$, then do a rotation by $\pi$ around the singularity, and then perform another
shift $z\to z+i\pi$. The asymptotic of the Bessel functions for large $|z|$ is given by \eqref{app:as}. The computation of the
monodromy is then straightforward:
\be
J_\nu(zw +i\pi w) = \cosh{\pi w} J_\nu(zw) - i \sinh{\pi w} N_\nu(zw), \nonumber \\
N_\nu(zw +i\pi w) = i \sinh{\pi w} J_\nu(zw)+ \cosh{\pi w} N_\nu(zw).
\ee
Thus, the monodromy matrix for the shift is:
\be
{\bf M}_{shift} = \left(
\begin{array}{cc}
\cosh{\pi w} & -i \sinh{\pi w} \\ i \sinh{\pi w} & \cosh{\pi w} 
\end{array}
\right).
\ee
Let us now calculate the monodromy for the rotation. This is an anti clock-wise rotation by $\pi$. The solutions transform as:
\be
J_\nu(z e^{\pi i}) = e^{\pi i \nu} J_\nu(z), \\ \nonumber
N_\nu(z e^{\pi i}) = e^{-\pi i \nu} N_\nu(z) + 2i \frac{\sin{\pi \nu}}{\tan{\pi \nu}} J_\nu(z).
\ee
The Bessel functions we need are for $\nu=0$. Then the last term in the transformation formula for $N_\nu$ has $0/0$ type singularity.
One has to understand the expression in the limiting sense. One then gets the following monodromy matrix:
\be
{\bf M}_{rot} = \left(
\begin{array}{cc}
1& 0 \\ 2i & 1 
\end{array}
\right).
\ee
The total monodromy matrix is ${\bf M}_h={\bf M}_{shift}{\bf M}_{rot}{\bf M}_{shift}$, and its trace is given by:
\be\label{mon-h-4}
\frac{1}{2}{\rm Tr}{\bf M}_h = \cosh{2\pi w} + \sinh{2\pi w}= e^{2\pi w}.
\ee

\bigskip
\noindent{\bf Singularity.} A closed cycle around the singularity is
a rotation by $4\pi$ in the $z$-plane. The monodromy one gets is:
\be\label{mon-s-4}
\frac{1}{2}{\rm Tr}{\bf M}_s = 1.
\ee

\bigskip
\noindent{\bf Infinity.} As before, the cycle around infinity is non-trivial, see Fig.\ref{fig:cont}.
One must first perform the shift $z\to z+i\pi$, then do a rotation, and then perform another
shift $z\to z+i\pi$. The monodromy matrix for the shift was calculated above. The computation of the monodromy for
rotation is also similar, but now the rotation is by angle $2\pi$ in the clock-wise direction. We get:
\be
{\bf M}_{rot} = \left(
\begin{array}{cc}
1& 0 \\ -4i & 1 
\end{array}
\right).
\ee
The total monodromy matrix is ${\bf M}_\infty={\bf M}_{shift}{\bf M}_{rot}{\bf M}_{shift}$, and its trace is given by:
\be\label{a2}
\frac{1}{2}{\rm Tr}{\bf M}_\infty = \cosh{2\pi w} - 2\sinh{2\pi w}.
\ee

\begin{figure}
\centering 
\epsfig{figure=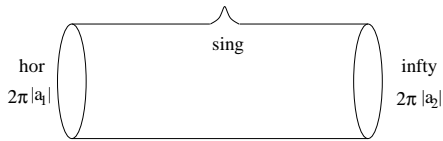, height=0.7in}
\caption{The $W$-surface in the high energy limit has two holes and one puncture.}
\label{fig:4-w}
\end{figure}

\bigskip
We now have all three monodromies, so we can immediately obtain the $W$-surface that arises in this approximation. The corresponding
wave equation is the hypergeometric one, with the parameters given by:
\be\label{index-4}
\cos{\pi a_1} = \cosh{2\pi w} + \sinh{2\pi w}, \qquad \cos{\pi a_2} = \cosh{2\pi w} - 2 \sinh{2\pi w}, 
\ee
and $a_3=0$. Let us also explicitly write down the parameters of the hypergeometric function:
\be\label{abg}
\alpha=\beta= \frac{1}{2} - \frac{a_1}{2} -\frac{a_2}{2}, \qquad \gamma = 1-a_1.
\ee
Thus, the $W$-surface in this approximation is the sphere with two holes: horizon plus infinity.
The object monodromy around which is
unity is the so-called puncture: it is a point of angle deficit $2\pi$. Thus, one gets
the $W$-surface shown in Fig.\ref{fig:4-w}.
%

Let us now discuss the quasi-normal modes. Once we have determined the wave equation to be the hypergeometric one it is 
straightforward to get the quasi-normal modes. They are zeros of the coefficient in front of the term in
\eqref{app:h} that is interpreted as the wave incoming at infinity. It is easy to see that such zeros can only come
from the poles of $\Gamma(\alpha)$ and $\Gamma(\beta)$.
The poles of the $\Gamma$-function are located at negative integers. Thus, we have an equation for the quasi-normal modes:
\be\label{qnm-4-poles}
1-a_1-a_2=-2\ n, 
\ee
This is quite similar to the condition (\ref{aaa}) we had in the case of BTZ black hole (notice that
in the present case the third monodromy $a_3$ vanishes).
Let us multiply this by $\pi$ and take the exponent. We find:
\be\label{qnm-geom}
\cos{\pi a_2} = - \cos{\pi a_1}.
\ee
One can now use the formula \eqref{index-4} to get:
\be
\label{log-3}
e^{4\pi w} = -3.
\ee 
So that the QNM frequencies fall into the set
\be
w_n=T_H\ln 3-2\pi T_H i (n+\frac{1}{2})~~.
\label{qnmf}
\ee
This is the $\log{3}$ behavior at high damping derived by a different but related method in \cite{Motl}. Let us also note
that, as in the examples considered before, the condition for QNM's has a geometrical interpretation. The equation
\eqref{qnm-geom} simply says that the sizes of holes are equal. The sign here is important. It can be thought of as
coming from the puncture that is inserted into the $W$-surface, see Fig.\ref{fig:4-w}.

\bigskip
\subsection{Reflection/transmission coefficients.}
In the limits discussed when the wave equation can be solved exactly
one can find not only the quasi-normal modes but also calculate explicitly the reflection
(or equivalently, the transmission) coefficient. These coefficients determine what is called the grey-body factor
which shows how the Hawking radiation deviates from the purely thermal spectrum.
The quasi-normal modes have standard interpretation as poles in the reflection amplitude.
The latter however contains more information than just the location of its poles.
In the low energy limit the reflection/transmission coefficients have been calculated in
\cite{Solod}. Here we focus on the high energy limit. The calculation of the grey-body factors
for BTZ black hole was done in \cite{Sen}.

The general solution to the hypergeometric equation is a combination of two functions:
\be
\label{phi}
\phi(y)=A\ F(\alpha,\beta;\gamma;y)+B\ y^{1-\gamma}F(\alpha-\gamma+1,\beta-\gamma+1;2-\gamma;y),
\ee
where in terms of the $y$ coordinate the asymptotic infinity is at $y=1$ while the horizon
is at $y=0$. Expanding the solution near $y=1$ and using equation (\ref{app:h}) and
(\ref{z=0}) we find that $\phi(y)$ is a combination of a constant function and 
$(1-y)^{\gamma-\alpha-\beta}$.  For parameters (\ref{abg}) we find that $\gamma-\alpha-\beta=a_2(\omega)$.
To identify in- and out-going modes let us consider large real frequencies. In this limit
$a_1(w)\sim a_2(w)\sim 2i w$. Thus, the mode $y^{1-\gamma}$ is out-going at the horizon, and
the mode $(1-y)^{\gamma-\alpha-\beta}$ is in-going 
at infinity. Assuming that a particle comes from the horizon, scatters and then goes to infinity
we expect only  out-going modes to appear at infinity. Thus, the in-going at infinity mode should vanish.
This gives a condition on constants $A$ and $B$ in (\ref{phi}):
\be
A \frac{\Gamma(\gamma)}{\Gamma (\alpha)\Gamma (\beta)}+
B \frac{\Gamma(2-\gamma)}{\Gamma (\alpha-\gamma+1)\Gamma (\beta-\gamma+1)}=0~~.
\label{AB}
\ee
Now we should look at the solution at the horizon, i.e. when $y\rightarrow 0$.
The solution (\ref{phi}) behaves there as follows
\be
\phi(y)=A+B \ y^{1-\gamma},
\label{BA}
\ee
where $A$ and $B$ are same coefficients as in (\ref{phi}) and (\ref{AB});
it is  a combination of the constant function and a mode $y^{1-\gamma}$. Since $\gamma=1-2i w$
the latter mode is out-going at the horizon while the constant function represents the in-going
mode. The reflection amplitude for a particle emitted form horizon is given by
the ratio, $S(w)\sim A/B$. Using the relation (\ref{AB}) and the definition (\ref{abg})
for the parameters of the hypergeometric function we find the expression for the reflection
amplitude
\be
S(w)\sim -\frac{\Gamma(1+a_1(w))}{\Gamma (1-a_1(w))}\left( 
\frac{\Gamma(\frac{1}{2}-\frac{a_1(w)}{2}-\frac{a_2(w)}{2})}{\Gamma(\frac{1}{2}+\frac{a_1(w)}{2}-\frac{a_2(w)}{2})}
\right)^2~~.
\label{reflection}
\ee
This can be compared to the expression found in \cite{Solod} in the low energy case.
Two expressions have a similar structure. Note, however, that the poles in (\ref{reflection}) are double poles,
while the reflection amplitude in the low energy case has only simple poles.
In the case of non-rotating BTZ black hole the poles are doubled as well
\cite{Sen} for zero angular momentum $l$. This degeneracy is removed when either rotation
or non-vanishing angular momentum is included.
The location of the poles in (\ref{reflection}) gives the QNM as expected. 
It would be of interest to compare \eqref{reflection} with the expression
found in \cite{Neitzke}. We shall not pursue this in the present paper.

\section{Effective theory}
\label{sec:effective}

We are now in a position to interpret all the results obtained. We have seen how two types of surfaces naturally
appear when one considers the BH wave equation. One surface is the Riemann surface of the tortoise map. The
wave equation is defined on this surface. The wave equation can be interpreted as the Fuchsian equation.
This interpretation gives rise to the $W$-surface that solves the equation. The $W$-surface encodes properties of the potential, and
it is this surface that is fundamental for the interpretation that we are going to propose.

As a step in the direction of an effective theory interpretation let us make the following observation. In all (solvable) examples
that we have considered the $W$-surface is topologically a sphere with 3 marked points, and we completely characterized
the surface by its monodromies (sizes of holes). One can obtain the same $W$-surfaces in the CFT context. Indeed,
the semi-classical limit of Liouville theory is intimately related with the uniformization and Riemann surfaces. This is
explained, for example, in the classical paper \cite{Zamolo} that proposed an answer for the Liouville theory 3-point function.
The basic idea of this relation is that in the semi-classical limit insertion of Liouville vertex operators modifies
the Liouville equation by introducing sources into it. The solution of the resulting Liouville equation can
be interpreted as giving a metric with a hole or conical singularity at the point of insertion of the vertex operator.
The precise identification is as follows. Consider ``heavy'' vertex operators $V_\alpha = e^{\alpha\phi(z)}$ with $\alpha=\eta/b$. 
Here $\eta$ is a parameter that survives in the semi-classical approximation, $\phi(z)$ is the Liouville field, 
and $b$ is the Liouville coupling constant. 
The semi-classical limit corresponds to $b\to 0$. For real $\eta<1/2$ such an operator will create a conical singularity
of the deficit angle $4\pi\eta$. For a general $\eta$ the monodromy around the ``object'' created is: $\cos{\pi a}$
where $a=1-2\eta$. 

Consider now the 3-point function of Liouville theory where the vertex operators that one inserts are such that
in the semi-classical limit the monodromies created would exactly match the monodromies of the $W$-surface. Our
observation is that the poles of such 3-point function are exactly the QNM of the corresponding BH. Let us demonstrate
this for the BTZ black hole. 
\bigskip
\subsection{Liouville 3-point function: BTZ black hole.}
We take the expression \eqref{app:3} for the 
3-point function and substitute into it the parameters:
\be
\alpha_1 = \frac{1}{2b}-\frac{iw}{2b}, \qquad \alpha_2=\frac{1-\nu}{2b}, \qquad \alpha_3=\frac{1}{2b}-\frac{il}{2b}.
\ee
Then the 3-point function takes the form:
\be\label{3-btz}
&{}&C(\alpha_1,\alpha_2,\alpha_3) \sim \\ \nonumber 
&{}& \frac{\Upsilon(\frac{1-iw}{b})\Upsilon(\frac{1-\nu}{b})\Upsilon(\frac{1-il}{b})}
{\Upsilon(\frac{1-\nu}{2b}-\frac{i(w+l)}{2b})
\Upsilon(\frac{1-\nu}{2b}-\frac{i(w-l)}{2b})
\Upsilon(\frac{1-\nu}{2b}+\frac{i(w-l)}{2b})
\Upsilon(\frac{1+\nu}{2b}-\frac{i(w+l)}{2b})}.
\ee
We work in the semi-classical approximation in which $Q\sim1/b$.
Taking into account the position \eqref{app:zeros} of zeros of the function $\Upsilon(x)$ we see that the
poles of \eqref{3-btz} occur in the following 4 series:
\be\label{3-poles-btz}
w=\pm l -2i (n+h), \qquad w=\pm l + 2i (n+h), \\ \nonumber
w=\pm l -2i(n+\tilde{h}), \qquad w=\pm l +2i(n+\tilde{h}).
\ee
Here everywhere $n\in\ZZ^{\geq 0}$ and we have introduced:
\be
\tilde{h}=\frac{1}{2}(1-\nu)=\frac{1}{2}(1-\sqrt{1+m^2}).
\ee
The first line in \eqref{3-poles-btz} is the correct set of QNM frequencies \eqref{qnm-btz} together
with their complex conjugates. The second line are similar frequencies in which the conformal dimension
$h$ is replaced by the conjugate conformal dimension $\tilde{h}$. These two sets 
do not overlap. Because $\tilde{h}=1-h$ the two sets of frequencies can be combined as follows:
\be\label{3-poles-btz'}
w=\pm l + 2i(n+h), \qquad  w=\pm l - 2i(n+h), \qquad n\in\ZZ.
\ee
If $l\not=0, m\not=0$ then there is (for a generic $m$) no zeros in the numerator and \eqref{3-poles-btz'} are indeed the poles of the
whole 3-point function. For a special case that $m$ is such that $h$ is a positive integer or half-integer, there is a zero 
in the numerator, and the two sets \eqref{3-poles-btz'} coincide. The 3-point function has to be 
understood in the generalized function sense. We shall not treat this case.

Thus, the poles of the 3-point function are the correct quasi-normal modes of the BTZ black hole (plus the complex conjugate frequencies),
together with a similar set of modes with the conformal dimension $h$ replaced by the conjugate dimension $\tilde{h}$. This has
the following explanation. We have two linearly independent solutions of the wave equation. In asymptotically AdS space, such
as the BTZ black hole, one of them grows and the other dies off at infinity. Quasi-normal modes are frequencies when there
is no growing mode (provided there is only the in-going mode at the horizon). However, there is another set of frequencies
for which there is no decaying mode. The 3-point function treats both linearly independent solutions on equal footing.
Thus, its poles contain both sets of frequencies .The fact 
that (\ref{3-btz}) contains also poles with positive real part might be related to the fact
we consider the Euclidean correlation function. In Minkowski version we would have to use the 
retarded correlation function as was suggested in \cite{BSS}. Then the poles with positive 
imaginary part as well as some unphysical poles may actually disappear. 

\bigskip
\subsection{Liouville 3-point function: Schwarzschild black hole.}
In the case of the  Schwarzschild black hole the parameters of the Liouville
vertex operators read:
\be
\alpha_1=\frac{1}{2b}-\frac{a_1}{2b},~~\alpha_2=\frac{1}{2b}-\frac{a_2}{2b},~~
\alpha_3=\frac{1}{2b}.
\label{part}
\ee
We see that one of the operators to be inserted is that corresponding to a puncture. The correct puncture vertex 
operator is: $\phi(z)e^{Q\phi(z)/2}=(\partial V_\alpha/\partial\alpha)_{\alpha=Q/2}$.
Thus, the 3-point function must be differentiated with respect to $\alpha_3$ before setting it to the value
\eqref{part}. This removes $\Upsilon(1/b)=0$ that would otherwise be present in the the numerator of \eqref{app:3}.
The 3-point function then takes the form
\be\label{3-sch}
&{}&C(\alpha_1,\alpha_2,\alpha_3) \sim \\ \nonumber 
&{}& \frac{\Upsilon(\frac{1-a_1}{b})\Upsilon(\frac{1-a_2}{b})}
{\Upsilon(\frac{1}{2b}-\frac{a_1}{2b}-\frac{a_2}{2b})
\Upsilon(\frac{1}{2b}-\frac{a_1}{2b}-\frac{a_2}{2b})
\Upsilon(\frac{1}{2b}+\frac{a_1}{2b}-\frac{a_2}{2b})
\Upsilon(\frac{1}{2b}-\frac{a_1}{2b}+\frac{a_2}{2b})}
\ee

The poles of the 3-point function \eqref{3-sch} fall into the following two sets:
\be
-\frac{a_1}{2}-\frac{a_2}{2}=\pm (n+\frac{1}{2}), \qquad \frac{a_1}{2}-\frac{a_2}{2}=\pm (n+\frac{1}{2}),
\label{sets}
\ee
where $n$ is non-negative integer. Notice that each pole appears twice in the
denominator of (\ref{3-sch}). It is easy to see that both sets satisfy \eqref{log-3}. 
The two sets of modes have the following interpretation: half of the frequencies are
the correct QNM's and their complex conjugates. 
Each monodromy $a_1$ and $a_2$ is a multi-sheet function of $w$, and for the other half of the
modes the monodromies are taken on a different sheet. Presumably, this other set corresponds to
``unphysical'' boundary conditions, such as purely outgoing
boundary conditions at the horizon and purely in-going conditions at infinity. We shall not
attempt to demonstrate this.

We conclude that
the Liouville 3-point function (\ref{3-sch}) reproduces the structure of the poles
of the reflection amplitude for Schwarzschild black hole we have calculated in section 4.3.
Both in the reflection amplitude and in the 3-point function \eqref{3-sch} the QNM's
appear as double poles. However, a precise 
relation between the reflection amplitude $S(w)$ given in section 4.3 and
the Liouville 3-point function just discussed is not clear at the moment. Presumably it should
be a generalization of the reflection property (noticed in \cite{Zamolo}) 
of Liouville 3-point function,
$C(Q-\alpha_1,\alpha_2,\alpha_3)=C(\alpha_1,\alpha_2,\alpha_3)S_L(i\alpha_1-iQ/2)$
where $S_L(P)$ is the Liouville reflection amplitude.

\bigskip
\subsection{The proposal.}
As we have demonstrated, the Liouville 3-point function \eqref{3-btz} does contain the BTZ quasi-normal modes as its 
poles. Similarly, the 3-point function \eqref{3-sch} contains the Schwarzschild QNM's in the high damping limit. 
We take this as a suggestion that the 3-point function describes the effective string propagating in the corresponding black hole
background. Probably the most non-trivial point in our construction is that it is the classical BH wave equation that
leads to the $W$-surface, and then the $W$-surface receives the interpretation of the effective string worldsheet.
The Liouville theory should be interpreted as the worldsheet theory. In the case considered, when we introduced
an ansatz \eqref{ansatz} to separate variables in the wave equation, the motion only occurs in the radial direction.
Correspondingly, there is only one field in the worldsheet CFT: the Liouville field playing the role of the
radial direction. 

For the case of the BTZ black hole there is yet another CFT: the boundary one. It is interesting to note that the Liouville 
conformal dimension of the vertex operators
considered, e.g. with $\alpha=(1-\nu)/2b$, does not coincide with the conformal dimension of the
corresponding operators in the boundary CFT. Thus, the Liouville CFT whose 3-point function is given by \eqref{3-btz} must be
interpreted as a worldsheet CFT, not the boundary one.

Thus, we are proposing that, the $W$-surface that the BH wave equation defines has the interpretation of the worldsheet of 
the effective string moving in the BH background. The theory on the worldsheet is presumable very complicated;
there is no reason to expect it to be a conformal field theory. However, 
in the limits in which the wave equation becomes solvable the theory simplifies. In cases we have
analyzed the wave equation reduces to the hypergeometric
equation and the resulting worldsheet theory the Liouville theory; the poles of the 3-point function 
are the quasi-normal modes. The fundamental role of the quasi-normal modes is that
they determine the poles of a correlator for the effective string. 

From the analysis that we have presented it is impossible to determine the Liouville 
coupling constant. In the semi-classical
approximation $b$ completely drops out of all the formulas. However, the Liouville theory that appears in the low energy 
limit can be further characterized, see \cite{Solod}. Indeed, 
the near horizon wave equation with the potential \eqref{pot-low} is the Liouville equation. This allows to
identify the Liouville coupling constant with the horizon radius measured in the units of string length: 
\be\label{coupl}
1/b=2r_+/\sqrt{\alpha'}.
\ee
The string length $\sqrt{\alpha'}$ remains an arbitrary parameter. 
The exact poles (at finite $b$) of 3-point function then can be considered as an $\alpha'$-modification of the semi-classical QNM.
The effective theory in the high energy limit can
probably be thought of as the Liouville theory which is perturbed. Indeed, removing the strictly near-horizon
restriction one introduces the term $e^{2z}$ to the potential. That perturbs the Liouville theory. The theory 
may start to flow, and become a new theory in the ultra-violet. This would be the theory describing the string
in the high energy limit. Our analysis shows that this theory is also related to the Liouville theory: the
string propagation is described by a particular 3-point function. However, the Liouville coupling constant may
flow, so the identification \eqref{coupl} may become invalid.

Let us now discuss the interpretation of various operators that one inserts in the 3-point function. Let us first discuss the case of BTZ and
low energy limit of Schwarzschild BH. In this cases one operator is
\be\label{vert-prop}
e^{(Q/2+(i/2b)(w/2\pi T_H))\phi}.
\ee 
This operator is inserted at a point that corresponds to the horizon. It can be interpreted as the vertex operator describing
the emission of a quantum with frequency $w$ from the horizon. Other operators only depend on the angular momentum 
of the wave, and not on the frequency. This is because
the potentials of the corresponding wave equations are growing at infinity and therefore do not admit propagating 
asymptotic modes. Therefore, one does not insert the vertex operator describing a propagating mode at infinity. Let us
now turn to the case of high energy approximation of 4d Schwarzschild. In this case, as we have discussed in the main text,
the approximation used is that of zooming on the near-singularity region. In doing this one cuts away both the
near-horizon and near-infinity regions. The resulting wave equation has propagating solutions in what 
replaced both the horizon and infinity. Thus, the vertex operators inserted at the horizon and infinity 
do depend on frequency. However, neither one of them is of the simple form \eqref{vert-prop}. To summarize, in all cases the $W$-surface describes
propagation: In the case of BTZ black hole a mode of frequency $w$ that appears at the horizon travels through the potential
and gets converted to a mode of conformal dimension $h$ at infinity. In the case of low energy limit of Schwarzschild one
again has a mode of frequency $w$ emitted from the horizon, and the $W$-surface describes its propagation through the
potential. In the high energy limit the $W$-surface that we have described is only the near-singularity part of the full
$W$-surface and describes motion of the effective string in the near-singularity region. Vertex operators inserted
at the horizon and infinity depend in a non-trivial way on frequency.

It is natural to speculate that this picture extends to BH's with more complicated Riemann surfaces. Thus, as
we have seen, the surface corresponding to the 5d Schwarzschild has 4 points. It is natural to expect that
there are limits in which the effective theory describing the string is well approximated by the Liouville
theory on this 4-punctured sphere. The insertions depend on the monodromies of the corresponding Riemann surface.

Thus, the resulting interpretation is that the $W$-surface describes string propagation. Note, however,
that the 3-point function is not equal to the transmission or scattering amplitude, even though all these
objects have partially overlapping sets of poles. This is not surprising; one needs to come up with further
rules on how the string amplitudes are to be extracted from the Liouville correlators. In the case of the
BTZ black hole there is a dual theory, and the full Hawking rate can be obtained by performing the
Fourier transform on the thermal 2-point function of the dual theory. 
The QNM then can be identified with the poles in 2-point function in the dual conformal theory, as was shown in \cite{BSS}.
No such rules have been proposed in our
case. Moreover, apart from the asymptotically AdS case, one does not expect a dual CFT to exist. To get
the physically interesting quantities one would have to do the worldsheet CFT calculation. It would
be of considerable interest to supplement the interpretation proposed in this paper with concrete
rules on how to extract physically interesting amplitudes from Liouville CFT. We leave development
of such rules for future work.

\section{Discussion}

In this paper we have introduced a notion of $Z$ and $W$-surfaces for the black hole wave equations. The $Z$-surface
arises as the image of the complex $r$-plane under the tortoise map. We have shown how this is in general a sphere
with a number of holes corresponding to horizons, plus two marked points corresponding to the singularity and infinity.
The black hole wave equation is the Fuchsian differential equation on the $Z$-surface. The ratio of two 
linearly independent solutions of this equation gives a map $w(z)$ from the $Z$-surface to the $W$-surface, which
is of the same topological type. The $W$-surface ``solves'' the wave equation, in that one can write down
an explicit solution if the map $w(z)$ (and its inverse) are known. The $Z$-surface does not know anything
about the potential. This information is encoded in the $W$-surface. 

For 4d Schwarzschild BH the
$Z$-surface is a sphere with 3 points. Thus, the hypergeometric equation can be used as an approximation
to the wave equation is some limits. We have analyzed two such limits: that of low and high energy. It is easy to
write down the quasi-normal modes of the resulting exactly solvable potential. In the limit of high energy this results in QNM's with 
$\log{3}$ as the real part of the frequency. We have explained that taking a limit in which the wave
equation becomes solvable amounts to zooming onto a part of the $W$-surface. Thus, the low energy limit
is the near horizon limit, and the high energy limit is the near singularity limit. In such limits the
$W$-surface is completely characterized by its monodromies, and is easy to determine.

For higher-dimensional BH's the $Z$-surface has more than 3 marked points. Thus, the hypergeometric equation
is no longer applicable. In the high energy limit $|w|\to\infty$ one zooms onto the region near the singularity. This
effectively cuts away all the holes
except the one corresponding to the horizon. The problem reduces to the already analyzed problem of 4d Schwarzschild. Thus,
this limit is insensitive to the increased complexity of the surface. It would be of interest to
find some other limits that lead to exactly solvable systems for which the presence of other
holes is important.

The main novelty of the present paper is our proposal that $Z,W$ Riemann surfaces signal the appearance of
an effective string. The motion of the string in the BH background is described by a theory on the
corresponding Riemann surface. This theory is complicated, there is no reason to expect that it is
conformal. However, in certain limits the theory simplifies, and can be described using the
Liouville CFT. Thus, our proposal is that the propagation of the string through the potential is described by
the Liouville 3-point function, where the operators to be inserted have to chosen as explained in the main text. 
We have shown how the poles of such 3-point function contain the BH quasi-normal modes. 

The description we gave is very general. It can be applied to any spherically symmetric BH. It would be of interest to
extend this description to more complicated BH's, such as rotating ones. Experience with the rotating BTZ BH suggests
that inclusion of rotation will simply make all the monodromies complex. The imaginary part will encode the
angular momentum. It would be very interesting to see whether this is indeed the case. It would also be important
to clarify the meaning of the moduli that start appearing for dimensions $n\geq 5$. In this case the surfaces
have more than 3 marked points. The number of arising moduli is $2n-8$. It would be interesting to give an
interpretation of this moduli in terms of the BH wave equation.

The stringy description we proposed is an effective one. It does not answer the question what are the
fundamental degrees of freedom of the BH. We are only saying that whatever the fundamental theory
describing the BH is, there must be the Liouville mode in it, so that the semi-classical properties
such as the Hawking emission are reproduced as sketched here. Thus, analysis of the BH wave equation
is not sufficient to determine how to describe the BH quantum mechanically. To find such a description
an additional input from some quantum gravity theory is necessary. However, the effective structure
we have found puts constraints on the fundamental theory. In particular, it is clear that,
whatever the fundamental theory is, it should have interpretation in CFT terms. In other words, the theory must
have a stringy interpretation. We believe that the structure unraveled
in this paper will be important in future searches for a fundamental theory describing black holes
quantum mechanically.

\section*{Acknowledgments} We would like to thank G. Arutyunov, V. Mukhanov  and I. Sachs for important discussions.
K.K. is grateful to the Institute for Theoretical Physics, Ludwig Maximilians University, Munich, and S.S. is grateful to 
the Albert Einstein Institute, Golm for hospitality while part of this work was carried out. S.S. thanks 
the KITP, UC at Santa Barbara, for hospitality  during initial
stages of this project. Research of S.S. is supported in part by the grant
DFG-SPP 1096 and by the National Science Foundation under Grant No. PHY99-07949.

\appendix
\section{Some facts on the hypergeometric equation}
\label{app:hyper}

The Fuchsian form of the hypergeometric equation is given by:
\be
T(z) = \frac{1-a_1^2}{z^2} + \frac{1-a_2^2}{(z-1)^2} +\frac{a_1^2+a_2^2-a_3^2-1}{z(z-1)}.
\ee
It can be reduced to the usual hypergeometric form:
\be
z(1-z) \frac{d^2 u}{dz^2} + [\gamma-(\alpha+\beta+1)z]\frac{d u}{dz} -\alpha\beta u = 0
\ee
where the parameters are related to those of the equation of the Fuchsian form by:
\be
a_1^2=(1-\gamma)^2, \qquad a_2^2=(\gamma-\alpha-\beta)^2, \qquad a_3^2 = (\alpha-\beta)^2.
\label{parameters}
\ee

As two linearly independent solutions around zero one usually takes:
\be
F(\alpha,\beta;\gamma;z), \qquad z^{1-\gamma} F(\alpha-\gamma+1,\beta-\gamma+1;2-\gamma;z).
\ee
It is easy to show that the monodromies around singular points, which are $0,1,\infty$, are given by:
\be
\frac{1}{2}{\rm Tr}{\bf M}_i = \cos{\pi a_i}.
\ee
The ratio of two linearly independent solutions maps the complex plane minus $0,1,\infty$ to a sphere
with 3 holes or conical singularities, depending on what $a_i$ are. Thus, if the trace of the monodromy
matrix is real and less than two one gets a conical singularity. If it is real and larger than two one gets a hole.
For complex trace the ``object'' on the $W$-surface is neither a hole or a conical singularity. In a sense,
it is a mixture of the two. 

Near $z=0$ the characteristic expansion is:
\be\label{z=0}
F(\alpha,\beta;\gamma;z) =1+\frac{\alpha\beta}{\gamma}z+\frac{\alpha (\alpha+1)\beta (\beta+1)}{\gamma (\gamma+1) 2}z^2+ ..
\ee

We also need the formula for the analytic continuation of the hypergeometric function from $z=0$ to $z=1$. One has:
\be\label{app:h}
F(\alpha,\beta;\gamma;z) = \frac{\Gamma(\gamma)\Gamma(\gamma-\alpha-\beta)}{\Gamma(\gamma-\alpha)\Gamma(\gamma-\beta)}
F(\alpha,\beta;\alpha+\beta-\gamma+1;1-z) + \\ \nonumber
(1-z)^{\gamma-\alpha-\beta} \frac{\Gamma(\gamma)\Gamma(\alpha+\beta-\gamma)}{\Gamma(\alpha)\Gamma(\beta)}
F(\gamma-\alpha,\gamma-\beta;\gamma-\alpha-\beta+1;1-z).
\ee

\section{Some facts on the Bessel functions}
\label{app:bessel}

The differential equation:
\be
u'' + \left( \beta^2 - \frac{4\nu^2-1}{4z^2} \right) u = 0
\ee
is solved in terms of Bessel functions via:
\be
u = \sqrt{z} J_{\nu}(\beta z).
\ee
The function $J_\nu(z)$ have a branch cut along the negative real axis. The values on opposite sides of the cut are related by:
\be\label{bessel-mon}
J_\nu(e^{i\pi k} z) = e^{i\pi k\nu} J_\nu(z).
\ee
Similarly, for the Bessel function of the second kind one has:
\be\label{sec-bessel-mon}
N_{\nu}(e^{i\pi k} z) = e^{-i\pi k\nu} N_\nu(z)+ 2i \frac{\sin{k\nu\pi}}{\tan{\nu\pi}} J_\nu(z).
\ee
The asymptotic of the Bessel functions is given by:
\be\label{app:as}
J_\nu(z) = \sqrt{\frac{2}{\pi z}} \cos{\left(z-\frac{\nu\pi}{2}-\frac{\pi}{4}\right)}+O(1/|z|), \\
N_\nu(z) = \sqrt{\frac{2}{\pi z}} \sin{\left(z-\frac{\nu\pi}{2}-\frac{\pi}{4}\right)}+O(1/|z|). \nonumber
\ee

\section{Some facts on the confluent hypergeometric function}
\label{app:confl}

The confluent hypergeometric equation is obtained from the usual hypergeometric equation when the singular point at $z=1$ is
taken to infinity. The equation one gets as the result of such limit is, when written in the Fuchsian form:
\be\label{whitt}
\frac{d^2 W}{dz^2} +\left( -\frac{1}{4} +\frac{\lambda}{z}+\frac{\frac{1}{4}-\mu^2}{z^2} \right)W = 0.
\ee
As two linearly independent solutions of this equation one can take Whittaker's functions: $W_{\lambda,\mu}(z), W_{-\lambda,\mu}(-z)$.

For Whittaker's functions $z=0$ is a branch point and $z=\infty$ is an essential singular point. 

For large values of $|z|$ we have the following asymptotic behavior:
\be\label{whitt-as}
W_{\lambda,\mu}(z) \sim e^{-z/2} z^\lambda.
\ee

\section{Liouville 3-point function}
\label{app:Liouv}

The expression for the Liouville 3-point function, as proposed in \cite{Zamolo}, is given by:
\be\label{app:3}
&{}&C(\alpha_1,\alpha_2,\alpha_3) = \left[ \pi \mu \gamma(b^2) b^{2-2b^2} \right]^{(Q-\sum \alpha_i)/b} \times \\ \nonumber
&{}&\frac{\Upsilon_0\Upsilon(2\alpha_1)\Upsilon(2\alpha_2)\Upsilon(2\alpha_3)}{\Upsilon(\alpha_1+\alpha_2+\alpha_3-Q)
\Upsilon(\alpha_1+\alpha_2-\alpha_3)\Upsilon(\alpha_2+\alpha_3-\alpha_1)\Upsilon(\alpha_3+\alpha_1-\alpha_2)}.
\ee
Here $\mu, b$ are the Liouville ``cosmological'' and the coupling constants correspondingly, $Q=b+1/b$, 
$\gamma(x)=\Gamma(x)/\Gamma(1-x)$, and $\Upsilon(x)$ is an entire function of $x$ with zeros located at:
\be\label{app:zeros}
x=-m/b-nb, \qquad x=Q+m/b + nb, \qquad m,n\in\ZZ^{\geq 0}.
\ee
An explicit expression for it can be found in \cite{Zamolo}. The quantity $\Upsilon_0$ is defined as:
\be
\Upsilon_0=\left(\frac{d\Upsilon(x)}{dx} \right)_{x=0}.
\ee

\end{document}